# CREDIT TO PIONEERING WORK ON CARBON NANOTUBES

**Eugene A. Katz**[1,2,*]

[1]Department of Solar Energy and Environmental Physics,

J. Blaustein Institutes for Desert Research, Ben-Gurion University of the

Negev, Sede Boqer Campus, 84990 Israel;

[1]Ilse-Katz Institute for Nanoscale Science and Technology,

Ben-Gurion University of the Negev, Beer Sheva 84105 Israel.

**Summary.** This letter gives a credit to a pioneering paper by A. M. Nesterenko et al (Izvestia Akademii Nauk SSSR, Met. 1982, [in Russian]) that is almost unknown to scientific community. On the basis of Transmission Electron Microscopy images and X-ray Ray Diffraction patterns of "carbon multi-layer tubular crystals" the authors suggested a model of nanotube structure formation and hypothesis on various chirality of carbon nanotubes.

---

[*]Corresponding author. Tel/Fax: + 972-86596736, E-mail address: keugene@bgumail.bgu.ac.il

The history of discovery of carbon nanotube (CNT) is repeatedly discussed [1]. Among other CNT pioneers, a credit is given to Russian physical chemists L. V. Radushkevich and V. M. Lukyanovich who, in 1952 in Russian Journal of Physical Chemistry, published clear Transmission Electron Microscopy (TEM) images of 50 nanometer diameter carbon tubes (Multi-Walled CNT in modern terminology ) synthesized by a thermocatalytical disproportionation of carbon monoxide [2].

However, one important pioneering work has not been given a credit as yet.

In 1982 (nine years before publication of the famous Iijima's paper [3] (1991)] the group of Ukrainian scientists published in Russian the results of chemical and structural characterization of carbon nanoparticles also produced by a thermocatalytical disproportionation of carbon monoxide [4].

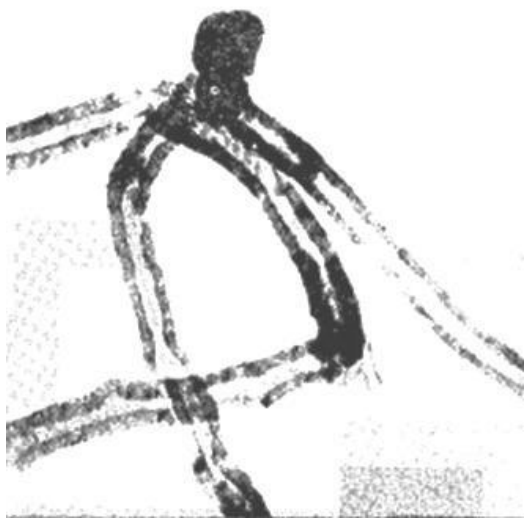

Fig. 1. Example of TEM images (X104,000) of carbon nanotubes with hollow channels and the catalyst particles connected to one end of the tubes published in [4] (reprinted by permission of Nauka Publishers).

The authors not only demonstrated TEM images of carbon nanotubes with hollow channels and the catalyst particles connected to one end of the tubes (Fig. 1) but analyzing the X-ray diffraction (XRD) data suggested that their "carbon multi-layer tubular crystals" were formed by rolling graphene layers into cylindrical tubes. It should be noted that similar conclusion was reached in a previous publication as well [5]. It differs this study from other early publications on manufacturing carbon filaments [6-8].

However, the authors of ref. 4 understood that such a rolling requires a circuit of hexagonal carbon nets of graphene into a cylinder without "seams". This was the first ever-published model of CNT structure formation. Furthermore (!), the authors formulated a conjecture on chirality of nanotubes. They speculated that during such rolling graphene layers into a cylinder, many different arrangements of graphene hexagonal nets are possible. They suggested two possibilities of such arrangements (Fig. 2): circular arrangement (armchair nanotube in the nowadays terminology) and a spiral, helical arrangement (chiral tube).

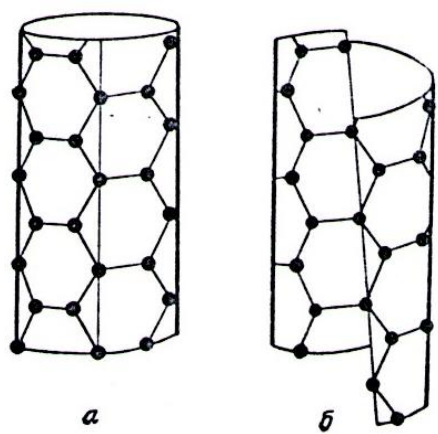

Fig. 2. Schematic representation of two different arrangements of graphene hexagonal nets (002) rolled into a cylinder, published in [4]: (a) armchair tube, (b) chiral tube (the helix pitch is equal to the corresponding lattice parameter of graphite). Rreprinted by permission of Nauka Publishers).

Unfortunately, to date, this article is almost unknown to scientific community. This pioneering work should get an international credit and be returned to the history of science.